\documentclass[11pt]{article}

\usepackage[utf8]{inputenc}
\usepackage[T1]{fontenc}
\usepackage[english]{babel}

\usepackage{amsmath, amssymb, amsthm}

\usepackage{graphicx}
\graphicspath{{figures/}}
\usepackage{float}
\usepackage{geometry}
\usepackage{tikz}
\usepackage{hyperref}
\hypersetup{
colorlinks=true, linkcolor=blue, citecolor=blue, urlcolor=blue, pdfauthor={Hector M. Trevino,
M.S.}, pdftitle={PTI: A Hierarchical Symbolic Color Addressing System} }

\usepackage{caption}
\usepackage{tocloft}
\usepackage[compact]{titlesec}
\titleformat*{\section}{\normalfont\large\bfseries}
\titleformat*{\subsection}{\normalfont\normalsize\bfseries}
\titleformat*{\subsubsection}{\normalfont\normalsize\itshape}
\usepackage[bottom]{footmisc}
\theoremstyle{plain}
\newtheorem{theorem}{Theorem}[section]
\newtheorem{corollary}{Corollary}[section]

\theoremstyle{definition}
\newtheorem{definition}{Definition}[section]
\newtheorem{property}{Property}[section]
\newtheorem{remark}{Remark}[section]

\newcommand{\Rot}{\mathrm{Rot}}
\newcommand{\ROL}{\mathrm{ROL}}

\usepackage{algorithm}
\usepackage{algpseudocode}

\usepackage{booktabs}
\usepackage{array}

\usepackage{csquotes}
\usepackage[backend=biber]{biblatex}
\usepackage{xcolor}
\definecolor{anchorBlack}{RGB}{0,0,0}
\definecolor{anchorBlue}{RGB}{0,0,255}
\definecolor{anchorGreen}{RGB}{0,255,0}
\definecolor{anchorCyan}{RGB}{0,255,255}
\definecolor{anchorRed}{RGB}{255,0,0}
\definecolor{anchorMagenta}{RGB}{255,0,255}
\definecolor{anchorYellow}{RGB}{255,255,0}
\definecolor{anchorWhite}{RGB}{255,255,255}
\definecolor{swatchBorder}{HTML}{999999}
\usepackage{datetime}
\usepackage{authblk}

\usepackage{orcidlink}

\title{PTI: A Hierarchical Symbolic Color Addressing System}
\author[1]{Hector M. Trevino, M.S.\,\orcidlink{0009-0005-4092-4704}}
\date{}

\begin{document}

\maketitle

\begin{abstract}

This paper introduces PTI (Persistent Traversal Identity), a hierarchical symbolic addressing
system for digital color derived from the recursive binary subdivision of the RGB color
cube. Standard hierarchical color processing systems organize colors via recursive space
partitioning but conventionally discard the traversal path once computation is complete. PTI
formalizes the preservation of that traversal path as a persistent symbolic identifier. The
resulting address is a finite sequence of eight-valued digits from $\{0,\ldots,7\}$, each
encoding the octant selected at one level of subdivision. Concatenated, they constitute
a lossless alternative representation of any 8-bit-per-channel RGB color.

PTI is structurally equivalent to the Morton (Z-order) code of a color interpreted
as a sequence rather than an integer, and its construction algorithm is the octree
traversal procedure of classical octree quantization, as established by Gervautz and
Purgathofer~\cite{gervautz1988simple}. PTI does not introduce a new color space, nor
does it claim to improve upon these foundational methods. Its contribution is a color
representation that is simultaneously machine-precise and hierarchically interpretable,
revealing a color's family lineage directly from its address structure. The same 8-digit
address decodes losslessly to RGB and reads compositionally as a named sequence of weighted
color directions.

PTI's well-defined prefix semantics support direct address operations suited to
color catalog organization, truncation-based image-processing routines, and structured
exploration of colors. The address structure invites further operations
beyond this. A simple one, successively rotating the address by one digit position, creates
structurally harmonious color palettes derived directly from the geometry of the RGB cube,
never leaving RGB.

\end{abstract}
\newpage

\section{Introduction} \label{sec:intro}

Every color within the RGB cube can be located through recursive subdivision.
Bisect the cube, select the octant containing the desired color, and repeat. Each
selection identifies a direction (toward Red, Cyan, Yellow) and reduces magnitude
by half at each step. These eight directions comprise the three additive primaries
(Red, Green, Blue), the three subtractive primaries (Cyan, Magenta, Yellow),
and the neutral extremes (Black, White).

Octree-based color quantization, introduced by Gervautz and
Purgathofer~\cite{gervautz1988simple}, organizes the RGB cube into a recursive tree. A
similar treatment underpins Morton (Z-order) codes~\cite{morton1966}, which interleave
the bits of spatial coordinates to produce hierarchically coherent linear indices.

In practice, both systems treat the traversal path as an ephemerally computed object to
build or index a structure. This paper proposes retaining the path as a color's identity,
its \textit{persistent traversal identity} (``PTI'').

Current color encoding formats like RGB-255 triplets, hexadecimal codes, and HSL/HSV specify
a color's position but not its traversal path. PTI simultaneously ensures
exact machine representation and hierarchical interpretability for practitioners. PTI retains
this path as a persistent identifier, bridging both requirements.

It formalizes this concept by representing each color through an ordered sequence of
octant indices from eight levels of RGB cube subdivision. Each digit signifies a weighted
contribution toward one of the anchor colors, halving in magnitude with each position. The
full address thus reveals the compositional path. PTI offers several benefits:
\begin{itemize}
\item \textbf{Lossless Encoding}: Encodes and decodes the exact 24-bit RGB value losslessly
in both directions.
\item \textbf{Dual Vocabulary}: Two-channel anchor digits carry both a subtractive name (Cyan,
Magenta, Yellow) and an additive RGB decomposition.
\item \textbf{Address Operations}: Supports direct operations on the address itself,
including additive complement and truncation, enabling applications such as catalog
organization.
\item \textbf{Native Palettes}: Derives harmonious color palettes directly from the
geometry of the RGB cube through address rotation, performed entirely on the address
sequence itself without decoding to RGB or leaving RGB.
\item \textbf{Structured Exploration}: Supports drill-down navigation across the full color
space through hierarchical address prefixes.
\end{itemize}

PTI builds on established hierarchical subdivision techniques used in spatial indexing and
color quantization but uniquely retains the traversal path as a persistent
object.

\section{Background and Related Work} \label{sec:background}

Morton codes, also known as Z-order codes or Z-order curve indices~\cite{morton1966}, linearize
multi-dimensional data while preserving spatial locality through bit interleaving. Given an
$n$-dimensional integer coordinate $(x_0, \ldots, x_{n-1})$, the Morton code is formed by
interleaving the binary representations of these coordinates from most significant to least
significant bits.

For instance, in three dimensions with each coordinate represented in $k$ bits, the Morton
code interleaves bits as:
\[ M = z_0 y_0 x_0 \, z_1 y_1 x_1 \ldots z_{k-1} y_{k-1} x_{k-1} \]
where subscripts denote bit positions. This results in a $3k$-bit value (24 bits for 8-bit
RGB) that defines a space-filling Z-curve, clustering spatially close points and enabling
efficient lookups.

\textbf{Example}: For the RGB color $(74,180,163)$, the binary representations are:
\[ R = 01001010, \quad G = 10110100, \quad B = 10100011. \]
Interleaving bits MSB-first yields:
\[
\underbrace{011}_{3}\,\underbrace{100}_{4}\,\underbrace{011}_{3}\,\underbrace{010}_{2}\,
\underbrace{100}_{4}\,\underbrace{010}_{2}\,\underbrace{101}_{5}\,\underbrace{001}_{1}
\]
Interpreted as octal digits, this becomes \texttt{34324251}, which is the PTI address.

Morton codes are widely used in spatial databases, octree traversal, and hierarchical
spatial data structures.

\subsection{Structural Equivalence with Morton Codes}

PTI addresses and Morton codes are computationally equivalent and share an intersection
of purposes, differing only in how bits are grouped. Given an RGB coordinate $(r, g, b)$
with 8-bit binary representations, the 24-bit Morton code is:
\[ M(r,g,b) = r_0 g_0 b_0 \; r_1 g_1 b_1 \;\cdots\; r_7 g_7 b_7. \]
The PTI address is derived by partitioning $M(r, g, b)$ into 3-bit groups and interpreting
each group as an octal digit:
\[ a_i = r_i \cdot 4 + g_i \cdot 2 + b_i, \quad i = 0, \ldots, 7. \]

\subsection{Semantic Distinction}

Morton codes and PTI addresses share the same locality-preserving structure, as PTI is
structurally equivalent to Morton encoding as demonstrated above. The difference lies in
grouping. Morton interleaves individual channel bits without weighting, whereas PTI
groups each bit-plane's three channel bits into a single weighted digit, so that digit
alone (rather than three separate bits scattered across the sequence) identifies a
complete, nameable direction.

\subsection{Octree Color Quantization}

An octree is a tree data structure where each internal node has eight children, representing
an eight-way partition of a three-dimensional space. When applied to the RGB color cube,
an octree recursively subdivides each dimension at its midpoint, creating eight sub-cubes
(octants) at each level.

Octree-based color quantization was introduced by Gervautz and
Purgathofer~\cite{gervautz1988simple}. They established a recursive RGB cube subdivision
framework for palette reduction. Their approach consists of bit-level traversal assigning
colors to octree nodes. This enables a merge-by-depth strategy for palette reduction.

Together with Morton codes, these octree contributions form the foundational framework of
hierarchical RGB subdivision that PTI builds upon. PTI does not advance the methods described
in these works. Instead, it adopts their established framework and introduces the concept of
the traversal path as a persistent, symbolic identifier independent of quantization.

\subsection{General Hierarchical Spatial Indexing}

Hierarchical spatial indexing and partitioning methods include octrees, quadtrees, k-d trees,
R-trees, Morton (Z-order) indexing, and bounding volume hierarchies (BVHs).
These techniques organize data through recursive subdivision, where the traversal path is
inherent to the tree structure but not retained as an independent identifier.

Color clustering and palette systems (such as k-means in RGB space and median-cut
quantization) also employ hierarchical structures but do not preserve traversal paths as
symbolic identifiers.

An analogous approach exists outside computer graphics: Geohash~\cite{niemeyer2008geohash}
encodes geographic coordinates using Z-order traversal, representing a location's recursive
subdivision as a persistent base-32 string. PTI does something similar in color space, retaining
the traversal path just as Geohash does for geographic data. Both
methods leverage hierarchical organization by maintaining the traversal path as a legible
identifier rather than discarding it post-navigation.

\section{Formal Specification} \label{sec:formal-spec}

\subsection{Mathematical Foundations}

\subsubsection{RGB Color Space} An 8-bit RGB color is represented by a triple $(r, g, b)$
where $r, g, b \in \{0, \ldots, 255\}$. The set of all such colors, denoted as
$\mathcal{C}_{24}$, forms a finite, discrete lattice consisting of $256^3 = 16{,}777{,}216$
integer points in $\mathbb{Z}^3$. This is not a continuous region in $\mathbb{R}^3$. Rather,
``the RGB cube'' refers to the bounding geometry of this lattice, not to a continuum of
colors. The extreme vertices of this lattice correspond to the three additive primaries,
the three subtractive primaries, and the two neutral extremes (Black and White).

\subsubsection{Binary Representation}
\begin{definition}[Binary Channel Representation] \label{def:binary}
For a channel value $v \in \{0,\ldots,255\}$, denote its 8-bit binary
representation (most significant bit first) by $(v_0, \ldots, v_7) \in
\{0,1\}^8$, where $v = \sum_{i=0}^{7} v_i \cdot 2^{7-i}$.
\end{definition}
The $i$-th bit plane of a color $(r, g, b)$ is the triple $(r_i, g_i,
b_i) \in \{0, 1\}^3$, where $i \in \{0, \ldots, 7\}$. Bit plane 0 is the most significant.

\subsubsection{Recursive Subdivision} The RGB cube can be recursively subdivided into
octants by halving each dimension. Each octant is identified by a selection tuple $(p, q,
s) \in \{0, 1\}^3$ indicating whether each channel falls in the lower (0) or upper (1)
half of the parent region.
\begin{definition}[Octant Index] \label{def:octant-index}
The \emph{octant index} of an octant identified by $(p, q, s) \in \{0,1\}^3$ is
the integer:
\[
I(p, q, s) = 4p + 2q + s \;\in\; \{0, 1, 2, 3, 4, 5, 6, 7\}.
\]
The positional weights $4, 2, 1$ correspond to the R, G, B axes respectively.
\end{definition}
This defines a bijection between the eight octants and the octal digits
$\{0,\ldots,7\}$. The eight vertices of the RGB cube, the \emph{anchor colors},
correspond to the octant indices shown below.

\subsubsection{Anchor Colors}

\begin{center}
\begin{tabular}{@{}llll@{}}
\toprule
\textbf{Color} & \textbf{RGB} & \textbf{Octant Index} & \textbf{Swatch} \\
\midrule
Black & $(0,0,0)$ & \texttt{0} & {\setlength{\fboxsep}{0pt}\fcolorbox{swatchBorder}{anchorBlack}{\phantom{\rule{1.5em}{1.2ex}}}} \\
Blue & $(0,0,255)$ & \texttt{1} & {\setlength{\fboxsep}{0pt}\fcolorbox{swatchBorder}{anchorBlue}{\phantom{\rule{1.5em}{1.2ex}}}} \\
Green & $(0,255,0)$ & \texttt{2} & {\setlength{\fboxsep}{0pt}\fcolorbox{swatchBorder}{anchorGreen}{\phantom{\rule{1.5em}{1.2ex}}}} \\
Cyan & $(0,255,255)$ & \texttt{3} & {\setlength{\fboxsep}{0pt}\fcolorbox{swatchBorder}{anchorCyan}{\phantom{\rule{1.5em}{1.2ex}}}} \\
Red & $(255,0,0)$ & \texttt{4} & {\setlength{\fboxsep}{0pt}\fcolorbox{swatchBorder}{anchorRed}{\phantom{\rule{1.5em}{1.2ex}}}} \\
Magenta & $(255,0,255)$ & \texttt{5} & {\setlength{\fboxsep}{0pt}\fcolorbox{swatchBorder}{anchorMagenta}{\phantom{\rule{1.5em}{1.2ex}}}} \\
Yellow & $(255,255,0)$ & \texttt{6} & {\setlength{\fboxsep}{0pt}\fcolorbox{swatchBorder}{anchorYellow}{\phantom{\rule{1.5em}{1.2ex}}}} \\
White & $(255,255,255)$ & \texttt{7} & {\setlength{\fboxsep}{0pt}\fcolorbox{swatchBorder}{anchorWhite}{\phantom{\rule{1.5em}{1.2ex}}}} \\
\bottomrule
\end{tabular}
\end{center}

\subsection{PTI Specification}

The recursive binary subdivision assigns a unique sequence of 8 total octant indices to
each color. This sequence is the PTI address. PTI addresses therefore comprise another
representation of $|\mathcal{C}_{24}| = 8^8 = 256^3 = 16{,}777{,}216$.

\subsubsection{PTI Address}
\begin{definition}[PTI Address] \label{def:pti-address}
Let $(r, g, b) \in \mathcal{C}_{24}$ with channels in 8-bit binary representation
per Definition~\ref{def:binary}. The \emph{PTI address of depth $d$}
($1 \leq d \leq 8$) is the finite sequence:
\[
a^{(d)} = (a_0, \ldots, a_{d-1}) \;\in\; \{0,\ldots,7\}^d,
\]
where each digit $a_i$ (for $i = 0, \ldots, d-1$) is the octant index
(Definition~\ref{def:octant-index}) of the $i$-th bit plane:
\[
a_i = 4 r_i + 2 g_i + b_i, \quad i = 0, \ldots, d-1.
\]
The \emph{full PTI address} of $(r,g,b)$ is its address at depth $d = 8$,
written as the concatenated octal string $a_0 a_1 \cdots a_7$.
\end{definition}

\subsubsection{Linear Algebraic Formulation of PTI} Let $\mathbf{w} = (4, 2, 1)^\top$ and $\mathbf{d} =
(2^7, 2^6, \ldots, 2^0)^\top$ be the channel weight vector and positional weight vector. For
a color $(r, g, b)$:
\begin{align*}
\mathbf{a} &= \mathbf{B}^\top \mathbf{w} \;\in\; \{0,\ldots,7\}^8
    \qquad \text{(PTI address)} \\
\mathbf{c} &= \mathbf{B}\,\mathbf{d} \;\in\; \{0,\ldots,255\}^3
    \qquad \text{(RGB reconstruction)}
\end{align*}
where $\mathbf{B} \in \{0,1\}^{3 \times 8}$ is the bit matrix of the color, with row $j \in
\{0,1,2\}$ corresponding to channel $R$, $G$, $B$ respectively, and $B_{j,i}$ the $i$-th bit
(MSB-first) of that channel. Equivalently, in closed form, where $c_j$ denotes the value of
channel $j$ (i.e., $c_0=r$, $c_1=g$, $c_2=b$):
\[
\mathbf{B} = \left( \left\lfloor \frac{c_j}{2^{7-i}} \right\rfloor \bmod 2 \right)_{\substack{j=0,1,2 \\ i=0,\ldots,7}} \in \{0,1\}^{3\times 8}.
\]
Written out explicitly:
\[
\mathbf{B} =
\begin{bmatrix}
r_0 & r_1 & r_2 & r_3 & r_4 & r_5 & r_6 & r_7 \\
g_0 & g_1 & g_2 & g_3 & g_4 & g_5 & g_6 & g_7 \\
b_0 & b_1 & b_2 & b_3 & b_4 & b_5 & b_6 & b_7
\end{bmatrix}
\]

The address $\mathbf{a}$ is obtained by applying the channel weight vector $\mathbf{w}$ to the
columns of $\mathbf{B}$, and $\mathbf{c}$ is obtained by applying the positional weight vector
$\mathbf{d}$ to the rows of $\mathbf{B}$. The PTI address and the RGB triple are therefore two
different views of the same 24-bit binary object, obtained by reading along perpendicular axes
of the matrix. The matrix $\mathbf{B}$ can be losslessly reconstructed from either $\mathbf{a}$
or $\mathbf{c}$ alone, establishing a bijection between $\mathcal{C}_{24}$ and the PTI address
space.

Although PTI's construction admits this linear algebraic formulation, its inverse is a
discrete decoding operation rather than a linear one, since the injectivity of the mapping
arises from the binary constraint on $\mathbf{B}$, not from a linear inverse over $\mathbb{R}$.
PTI-Decode (Section~\ref{sec:pti-decode}) performs this inverse through bit extraction rather
than matrix inversion.

\subsection{PTI Algorithms}

\subsubsection{PTI-Encode} \label{sec:pti-encode} The PTI address of a color $(r, g, b)$ is
constructed by reading the three channel bit sequences in parallel, one bit position at a
time from MSB to LSB, and mapping each position to an octal digit.

\begin{algorithm}
\caption{PTI-Encode: RGB to PTI address}
\label{alg:pti-encode}
\begin{algorithmic}[1]
\Require Color $(r, g, b)$ with each channel in $\{0,\ldots,255\}$
\Ensure Full PTI address $(a_0, \ldots, a_7) \in \{0,\ldots,7\}^8$
\For{$i \gets 0$ \textbf{to} $7$}
\State $r_i \gets \lfloor r \;/\; 2^{7-i} \rfloor \bmod 2$
\Comment{$i$-th bit of $r$, MSB first}
\State $g_i \gets \lfloor g \;/\; 2^{7-i} \rfloor \bmod 2$
\Comment{$i$-th bit of $g$, MSB first}
\State $b_i \gets \lfloor b \;/\; 2^{7-i} \rfloor \bmod 2$
\Comment{$i$-th bit of $b$, MSB first}
\State $a_i \gets 4 r_i + 2 g_i + b_i$
\Comment{combine bits into the depth-$i$ PTI digit}
\EndFor
\State \Return $(a_0, \ldots, a_7)$
\end{algorithmic}
\end{algorithm}

\subsubsection{PTI-Decode} \label{sec:pti-decode} Reconstructing $(r, g, b)$ from a PTI address involves
recovering the bits one digit at a time, MSB to LSB, and summing them by bit weight in succession.

\[ r = \sum_{i=0}^{7} \left\lfloor \frac{a_i}{4} \right\rfloor \cdot 2^{7-i}, \quad g = \sum_{i=0}^{7} \left\lfloor
\frac{a_i \bmod 4}{2} \right\rfloor \cdot 2^{7-i}, \quad b = \sum_{i=0}^{7} (a_i \bmod 2) \cdot 2^{7-i}. \]

\begin{algorithm}
\caption{PTI-Decode: PTI address to RGB}
\label{alg:pti-decode}
\begin{algorithmic}[1]
\Require Full PTI address $(a_0, \ldots, a_7) \in \{0,\ldots,7\}^8$
\Ensure Color $(r, g, b) \in \{0,\ldots,255\}^3$
\State $r \gets 0$; \quad $g \gets 0$; \quad $b \gets 0$
\For{$i \gets 0$ \textbf{to} $7$}
\State $t \gets a_i$
\State $r \gets r + \lfloor t \;/\; 4 \rfloor \cdot 2^{7-i}$
\Comment{high bit of $t$ reconstructs R}
\State $g \gets g + \lfloor (t \bmod 4) \;/\; 2 \rfloor \cdot 2^{7-i}$
\Comment{middle bit reconstructs G}
\State $b \gets b + (t \bmod 2) \cdot 2^{7-i}$
\Comment{low bit reconstructs B}
\EndFor
\State \Return $(r, g, b)$
\end{algorithmic}
\end{algorithm}

PTI-Encode and PTI-Decode are exact inverses. Encoding a color and decoding the result
recovers the original color exactly, with no loss of information. A reference
implementation is available at \url{https://github.com/metrigram/pti}.

\subsubsection{Worked Example}

Consider $(r, g, b) = (255, 170, 0)$, a vibrant orange-yellow:
\begin{align*}
r &= 255 = (11111111)_2 \\
g &= 170 = (10101010)_2 \\
b &= 0 = (00000000)_2
\end{align*}

Applying Definition~\ref{def:pti-address} at each level:

\begin{center}
\begin{tabular}{@{}c@{\hspace{1.4em}}ccc@{\hspace{1.4em}}c@{}}
\toprule
\textbf{Level }$\boldsymbol{i}$ & $\boldsymbol{r_i}$ & $\boldsymbol{g_i}$ & $\boldsymbol{b_i}$ &
$\boldsymbol{a_i = 4r_i + 2g_i + b_i}$ \\
\midrule
0 & 1 & 1 & 0 & 6 \\
1 & 1 & 0 & 0 & 4 \\
2 & 1 & 1 & 0 & 6 \\
3 & 1 & 0 & 0 & 4 \\
4 & 1 & 1 & 0 & 6 \\
5 & 1 & 0 & 0 & 4 \\
6 & 1 & 1 & 0 & 6 \\
7 & 1 & 0 & 0 & 4 \\
\bottomrule
\end{tabular}
\end{center}

The full PTI address is \texttt{64646464}. The alternating pattern reflects the
alternating bits of $g = 170 = (10101010)_2$ against the constant all-1 red
channel and all-0 blue channel.

Inversely, to decode the PTI address \texttt{64646464} back to RGB, PTI-Decode is applied.
Each digit $a_i$ is split into its three bits, reconstructing the original channel bits:

$6 \to (1,1,0)$, $4 \to (1,0,0)$, etc. Reassembling the bits into the matrix $\mathbf{B}$
and summing each row by weight yields:

\begin{center}
\begin{tabular}{@{}l@{\hspace{1.4em}}cccccccc@{\hspace{1.4em}}c@{}}
\toprule
\textbf{Digit }$\boldsymbol{a_i}$ & 6 & 4 & 6 & 4 & 6 & 4 & 6 & 4 & \\
\textbf{Weight }$\boldsymbol{2^{7-i}}$ & 128 & 64 & 32 & 16 & 8 & 4 & 2 & 1 & \\
\midrule
$\boldsymbol{r_i}$ & 1 & 1 & 1 & 1 & 1 & 1 & 1 & 1 & $R=255$ \\
$\boldsymbol{g_i}$ & 1 & 0 & 1 & 0 & 1 & 0 & 1 & 0 & $G=170$ \\
$\boldsymbol{b_i}$ & 0 & 0 & 0 & 0 & 0 & 0 & 0 & 0 & $B=0$ \\
\bottomrule
\end{tabular}
\end{center}

\section{Fundamental Properties} \label{sec:properties}

\subsection{Gray Colors in PTI}

In the RGB color space, the neutral axis runs from Black $(0, 0, 0)$ to White $(255, 255,
255)$, encompassing all shades of gray where $R = G = B$. In PTI, this neutral axis is
characterized by addresses that use only the digits 0 (Black) and 7 (White).

\begin{property}[Gray Address Characterization] \label{prop:grays} A PTI address represents
a gray color if and only if every digit is either 0 or 7. \end{property}

\begin{proof} A color is gray when $R = G = B$, implying $r_i = g_i = b_i$ for each bit
position $i$. The corresponding PTI digit at position $i$ is given by $a_i = 4r_i + 2g_i +
b_i$. If all channel bits are zero ($r_i = g_i = b_i = 0$), then $a_i = 0$; if all are one
($r_i = g_i = b_i = 1$), then $a_i = 7$. No other digit can occur when the three channel
bits match. Conversely, digits 0 and 7 enforce $r_i = g_i = b_i$ for all $i$. \end{proof}

There are exactly 256 such addresses, one per shade of gray, confirming a bijective mapping
on the neutral axis. The entire neutral axis is encoded using only two anchor directions
(0 and 7) in PTI, excluding all six chromatic anchors (1, 2, 3, 4, 5, 6), which require
at least one non-gray digit:
\begin{itemize}
\item RGB $(128, 128, 128)$ corresponds to the binary value $10000000$ per channel.
In PTI, this is represented as \texttt{70000000}.
\item RGB $(192, 192, 192)$ corresponds to the binary value $11000000$ per channel. In PTI,
this is represented as \texttt{77000000}.
\end{itemize}

Both examples demonstrate that gray addresses use only digits 0 and 7, upholding
Property~\ref{prop:grays}.

\subsection{Canonical Resolution Levels} \label{sec:canonical-depths}

PTI depths $\{1, 2, 4, 8\}$ are canonical because they divide 8 evenly. At these depths,
the periodic-prefix representative tiles perfectly into an 8-digit address.

\begin{definition}[Canonical Depth] \label{def:canonical-depth} A depth $d \in \{1, 2, 4, 8\}$ is canonical if it
divides 8. The periodic representative of any $d$-digit prefix $p_0 \cdots p_{d-1}$ is formed
by repeating the prefix $8/d$ times. \end{definition}

Truncation can occur at any depth $\{1, \ldots, 7\}$, but only canonical depths
produce structurally regular periodic representatives that tile evenly across an
address. Non-canonical depths (3, 5, 6, 7) produce truncated prefixes that are not tileable
across a full address.

\subsubsection{Canonical Depths and Colors} The table below summarizes the representative-color
count, tiling factor, and bit resolution at each canonical depth:
\begin{center}
\begin{tabular}{@{}ccccl@{}}
\toprule
\textbf{Depth} & \textbf{Representative Colors} & \textbf{Tile} & \textbf{Bits/Channel} & \textbf{Description} \\
\midrule
1 & 8 & $\times 8$ & 1 & Anchor colors \\
2 & 64 & $\times 4$ & 2 & Minimal expressive color map \\
4 & 4{,}096 & $\times 2$ & 4 & Analog of CSS \texttt{\#RGB} \\
8 & 16{,}777{,}216 & $\times 1$ & 8 & Full 24-bit color \\
\bottomrule
\end{tabular}
\end{center}

These depths form a nested hierarchy under prefix inclusion:
\[ \mathcal{S}_1 \subseteq \mathcal{S}_2 \subseteq \mathcal{S}_4 \subseteq \mathcal{S}_8, \]
where $\mathcal{S}_d$ denotes the set of depth-$d$ periodic representatives.

\subsubsection{Coverage by Prefix Weight}

The first PTI digit alone resolves 50\% of each color channel's value by determining
which half of the 0--255 range the true value falls into. Each subsequent digit further
halves the remaining uncertainty, converging to the exact value at full depth. The
coverage below shows the fraction of the channel range resolved at each of the canonical
depths.

\begin{center}
\begin{tabular}{@{}ccr@{}}
\toprule
\textbf{Depth} & \textbf{Coverage}$^\dagger$ & \textbf{Max channel error} \\
\midrule
1 & 50.00\% & 128 units \\
2 & 75.00\% & 64 units \\
4 & 93.75\% & 16 units \\
8 & 100\% (exact) & 0 units \\
\bottomrule
\multicolumn{3}{l}{{\footnotesize $^\dagger$Percentages computed relative to $2^8 = 256$, a known off-by-one in binary}} \\
\multicolumn{3}{l}{{\footnotesize representations. Exact at $d=8$ by direct identification.}} \\
\end{tabular}
\end{center}

\begin{remark}[Compact Address Notation] \label{rem:compact} At canonical depth $d$,
the periodic representative can be denoted by its $d$-digit prefix alone, similar to CSS
\#RGB shorthand. For example, \texttt{4} represents $44444444$ (Red), \texttt{34}
represents $34343434$ (Cyan-Red), and \texttt{3432} represents $34323432$. This compact
form applies exclusively to periodic representatives (Section~\ref{sec:canonical-depths}).
\end{remark}

\definecolor{swatchD1}{RGB}{0,255,255}
\definecolor{swatchD2}{RGB}{85,170,170}
\definecolor{swatchD4}{RGB}{68,187,170}
\definecolor{swatchD8}{RGB}{74,180,163}

\textbf{Example Across Canonical Depths}: Consider the address \texttt{34324251}, which
decodes to RGB $(74, 180, 163)$. The address can be examined at all canonical depths:
\begin{center}
\begin{tabular}{@{}clllc@{}}
\toprule
\textbf{Depth} & \textbf{Prefix} & \textbf{Periodic rep} & \textbf{Reading} & \textbf{Swatch} \\
\midrule
1 & \texttt{3} & \texttt{33333333} = RGB$(0,255,255)$ & Cyan family & {\setlength{\fboxsep}{0pt}\fcolorbox{swatchBorder}{swatchD1}{\phantom{\rule{1.5em}{1.2ex}}}} \\
2 & \texttt{34} & \texttt{34343434} = RGB$(85,170,170)$ & Cyan$\to$Red & {\setlength{\fboxsep}{0pt}\fcolorbox{swatchBorder}{swatchD2}{\phantom{\rule{1.5em}{1.2ex}}}} \\
4 & \texttt{3432} & \texttt{34323432} = RGB$(68,187,170)$ & Cyan$\to$Red, Cyan$\to$Green & {\setlength{\fboxsep}{0pt}\fcolorbox{swatchBorder}{swatchD4}{\phantom{\rule{1.5em}{1.2ex}}}} \\
8 & \texttt{34324251} & RGB$(74,180,163)$ & exact color & {\setlength{\fboxsep}{0pt}\fcolorbox{swatchBorder}{swatchD8}{\phantom{\rule{1.5em}{1.2ex}}}} \\
\bottomrule
\end{tabular}
\end{center}

\begin{theorem}[Equivalence with CSS Shorthand Hex] \label{thm:css-equiv} The set of colors
encoded by PTI depth-4 periodic representatives matches the set of colors encoded by CSS
3-digit shorthand hex (\texttt{\#RGB})~\cite{w3c_css_color3}. Both represent the same subset
of the full $\mathcal{C}_{24}$ color space, even as different color encodings. \end{theorem}

A formal proof of the above theorem is provided in Appendix~\ref{app:proof-css-equiv}.

\subsection{Compositional Interpretation of PTI Addresses}

PTI addresses encode a color through a sequence of eight digits, each representing a
directional contribution in the RGB color space with decreasing magnitudes. The full color
is determined by summing these contributions:
\[
(r, g, b) = \sum_{i=0}^{7}
\bigl(\lfloor a_i/4\rfloor,\;
\lfloor(a_i \bmod 4)/2\rfloor,\;
a_i \bmod 2\bigr)
\cdot 2^{7-i}
\]
Each digit $a_i$ at position $i$ specifies a direction and magnitude, contributing to the
final color. For example, an address like \texttt{34324251} encodes a hierarchical path
through the RGB cube.

PTI addresses are readable at multiple canonical depths (1, 2, 4, and 8), each level adding
more detail while maintaining structural consistency. At any depth $d$, reading an address
requires identifying which of eight anchor colors corresponds to each digit. This consistent
vocabulary facilitates comprehension across levels. For instance, \texttt{3} captures the dominant
direction at depth-1 (Cyan family), \texttt{34} places the color in the Cyan-Red family at depth-2,
and \texttt{3432} further refines it to a Cyan-Green sub-family at depth-4. This hierarchical
composition allows for an easier mental bridge between PTI addresses and their corresponding
visual colors.

\subsubsection{Dual Vocabulary} Each PTI digit names one of the eight familiar anchor colors:
Red, Green, and Blue at the single-channel corners; Cyan, Magenta, and Yellow at the
two-channel corners between them; Black and White at the two extremes. The two-channel
anchors are not a separate system layered into PTI. They arise directly from the same
three-bit digit, as each is simply the sum of its two component channel digits:
\begin{itemize}
\item Cyan ($3$) = Green ($2$) + Blue ($1$)
\item Magenta ($5$) = Red ($4$) + Blue ($1$)
\item Yellow ($6$) = Red ($4$) + Green ($2$)
\end{itemize}
A practitioner needs nothing new here. The two-channel anchors are simply combinations of the
same familiar colors already used for the single-channel anchors. Although PTI addresses do
encode CMY anchors, this does not imply that PTI operates in a CMYK color space. The underlying
computation remains purely additive RGB.

\begin{center}
\begin{tabular}{@{}lllllrrrr@{}}
\toprule
\textbf{Pos.} & \textbf{Digit} & \textbf{Direction} & \textbf{Ch.} & \textbf{Swatch} & \textbf{Wt.}
  & $\Delta R$ & $\Delta G$ & $\Delta B$ \\
\midrule
1 & 3 & Cyan & 2 & {\setlength{\fboxsep}{0pt}\fcolorbox{swatchBorder}{anchorCyan}{\phantom{\rule{1.5em}{1.2ex}}}} & 128 & 0 & +128 & +128 \\
2 & 4 & Red & 1 & {\setlength{\fboxsep}{0pt}\fcolorbox{swatchBorder}{anchorRed}{\phantom{\rule{1.5em}{1.2ex}}}} & 64 & +64 & 0 & 0 \\
3 & 3 & Cyan & 2 & {\setlength{\fboxsep}{0pt}\fcolorbox{swatchBorder}{anchorCyan}{\phantom{\rule{1.5em}{1.2ex}}}} & 32 & 0 & +32 & +32 \\
4 & 2 & Green & 1 & {\setlength{\fboxsep}{0pt}\fcolorbox{swatchBorder}{anchorGreen}{\phantom{\rule{1.5em}{1.2ex}}}} & 16 & 0 & +16 & 0 \\
5 & 4 & Red & 1 & {\setlength{\fboxsep}{0pt}\fcolorbox{swatchBorder}{anchorRed}{\phantom{\rule{1.5em}{1.2ex}}}} & 8 & +8 & 0 & 0 \\
6 & 2 & Green & 1 & {\setlength{\fboxsep}{0pt}\fcolorbox{swatchBorder}{anchorGreen}{\phantom{\rule{1.5em}{1.2ex}}}} & 4 & 0 & +4 & 0 \\
7 & 5 & Magenta & 2 & {\setlength{\fboxsep}{0pt}\fcolorbox{swatchBorder}{anchorMagenta}{\phantom{\rule{1.5em}{1.2ex}}}} & 2 & +2 & 0 & +2 \\
8 & 1 & Blue & 1 & {\setlength{\fboxsep}{0pt}\fcolorbox{swatchBorder}{anchorBlue}{\phantom{\rule{1.5em}{1.2ex}}}} & 1 & 0 & 0 & +1 \\
\midrule
  & & \textbf{Total} & & & & \textbf{74} & \textbf{180} & \textbf{163} \\
\bottomrule
\end{tabular}
\end{center}

The address \texttt{34324251} illustrates this. Each digit provides a weighted RGB
contribution, with two-channel digits 3 and 5 involving two channels at once.

\subsubsection{Hue Sector Identification} 
\begin{figure}[H] \centering
\includegraphics[width=0.6\textwidth]{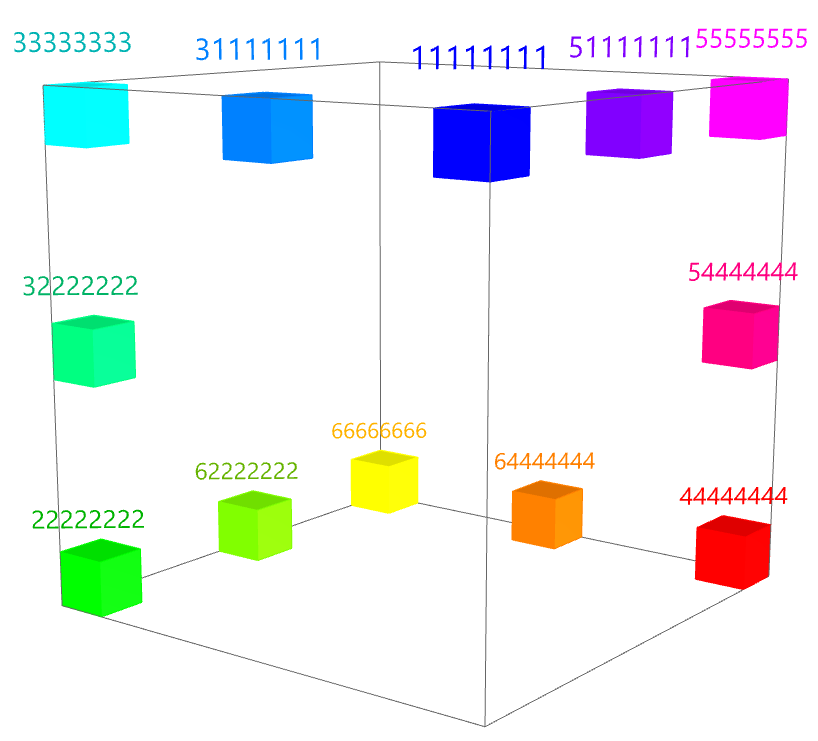}
\caption{Hues sampled at 30-degree intervals around the HSL color wheel, each cube labeled
with its PTI address. The anchor hues display a uniform, single-digit-repeated address,
while the midpoint hues show a mixed address.}
\label{fig:hue-sector}
\end{figure}

The six chromatic anchor colors sit at even 60-degree
intervals around the standard color wheel. Figure~\ref{fig:hue-sector} samples hues at
30-degree intervals (half the anchor spacing) to compare two cases:
\begin{itemize}
\item The six anchor hues decode to a uniform, single-digit-repeated PTI address.
\item The six midpoint hues between them decode to a mixed address, but not an arbitrary
one. Each splits cleanly in half, with the first digit alone accounting for half of the
color's composition (see the coverage discussion in Section~\ref{sec:canonical-depths}),
and a single repeated digit accounting for the other half across the remaining seven
digits.
\end{itemize}

\textit{Note: RGB cube renderings throughout this paper serve as stylistic aids to
provide spatial context, not as scaled geometric projections.}

Together, these properties reveal the structural value of PTI's encoding. Gray colors are
immediately intuitive. Every shade of gray is simply a mix of Black (0)
and White (7). The hue-sector midpoints reveal a similarly elegant structure. The address is
comprised of exactly two anchor digits, split cleanly half and half. Beyond these
individual properties, PTI provides practical utility as both a machine-precise and
hierarchically interpretable encoding, where deeper readings build upon simpler ones.

\section{Address Operations} \label{sec:address-ops}

\subsection{Additive Complement in PTI}

For any color $c = (R, G, B)$, its additive complement is defined as $c' = (255 - R, 255 -
G, 255 - B)$. In PTI addressing, the complement of a color with address $a = (a_0,
\ldots, a_7)$ is obtained by transforming each digit $t$ to $7 - t$.

\begin{property}[Additive Complement in PTI] \label{prop:complement} The PTI address of
the additive complement is derived by replacing each digit $t$ with $7 - t$. This operation
can be expressed as:
\[ (a'_0, \ldots, a'_7) = (7 - a_0, \ldots, 7 - a_7) \]
This transformation flips all bits in each channel, ensuring that the sum of corresponding
channels of $c$ and $c'$ equals white, $(255, 255, 255)$. \end{property}

\begin{proof}
Subtracting any 8-bit value from $255 = (11111111)_2$ flips all bits. At each position $i$,
if $(r_i, g_i, b_i)$ are the channel bits of $c$, the complement has bits
$(1-r_i,\,1-g_i,\,1-b_i)$. The PTI digit transforms as:
\[
4(1-r_i) + 2(1-g_i) + (1-b_i) \;=\; 7 - (4r_i + 2g_i + b_i) \;=\; 7 - a_i.
\]
This holds at every position and for every color without exception.
\end{proof}

\subsubsection{Complement Pairs Among Anchors} The table below pairs each anchor digit with its
additive complement:

\begin{center}
\begin{tabular}{@{}llcll@{}}
\toprule
\textbf{Digit} & \textbf{Anchor Color} & \textbf{Complement Pair} & \textbf{Digit} & \textbf{Anchor Color} \\
\midrule
0 & Black & $\leftrightarrow$ & 7 & White \\
1 & Blue & $\leftrightarrow$ & 6 & Yellow \\
2 & Green & $\leftrightarrow$ & 5 & Magenta \\
3 & Cyan & $\leftrightarrow$ & 4 & Red \\
\bottomrule
\end{tabular}
\end{center}

\textbf{Worked Example Beyond the Anchors}: The anchor pairs mentioned above are a simple case,
each anchor's full address consisting of a single repeated digit. Black's \texttt{00000000}
complements to White's \texttt{77777777}, since $(7,7,7,7,7,7,7,7) - (0,0,0,0,0,0,0,0) = (7,7,7,7,7,7,7,7)$.
Applying the same subtraction to this paper's running example, address \texttt{34324251} (RGB$(74, 180, 163)$):
\[
(7,7,7,7,7,7,7,7) - (3,4,3,2,4,2,5,1) = (4,3,4,5,3,5,2,6),
\]
yielding \texttt{43453526}, which decodes to RGB$(181, 75, 92)$.
This is exactly $(255-74,\,255-180,\,255-163)$. The property holds for any address, not just the
eight anchors.

\begin{remark}[Additive vs.\ Hue Complement] \label{rem:complement-hsl}
The operation $t \to 7 - t$ computes the additive complement directly. This differs
from the hue complement, which rotates the hue angle by $180^\circ$ while preserving
saturation and brightness. PTI natively computes the additive complement exactly and for
all colors. Computing the hue complement requires stepping outside the address into HSL
directly. Currently, there is no known direct transformation of the digits for the hue
complement.
\end{remark}

\subsection{Truncating Addresses} \label{sec:truncating}

A PTI address of depth $d$ (where $d < 8$) is obtained by retaining only the first $d$ digits
of an 8-digit full PTI address, representing a unique sub-cube within the RGB color space.

\begin{definition}[Truncated PTI Address] \label{def:truncated-address} The truncated PTI address of depth $d$ for a color $c$, denoted
$\text{prefix}^{(d)}(c) = (a_0, \ldots, a_{d-1})$, identifies the specific sub-cube containing
$c$. This sub-cube has an edge length of $256/2^d$ per channel and contains $(256/2^d)^3$
RGB colors sharing this prefix. \end{definition}

\textbf{Example}: The address \texttt{34324251} truncates to \texttt{3432} at depth 4 and
to \texttt{34} at depth 2. The depth-4 prefix \texttt{3432} corresponds to a sub-cube
of edge length 16 per channel, containing 4096 colors. The depth-2 prefix \texttt{34}
corresponds to a larger sub-cube of edge length 64, containing 262144 colors.

\begin{property}[Prefix Inclusion] \label{prop:prefix-inclusion} If two colors $c_1$ and $c_2$ share the same truncated PTI address of depth
$d$, they lie within the same sub-cube. This sub-cube has an edge length of $256/2^d$
per channel and contains $(256/2^d)^3$ RGB colors. \end{property}

Truncating a PTI address reduces color precision to $d$ bits per channel, mapping it to
a unique depth-$d$ sub-cube. All colors within this sub-cube share the same truncated address.

\subsubsection{Representative Color} For $d < 8$, a depth-$d$ address identifies a sub-cube
rather than a single color. At $d = 8$, the address uniquely defines both a sub-cube and a
color, requiring a representative point:
\begin{enumerate}
\item \textbf{Periodic-prefix convention}: The periodic representative
(Section~\ref{sec:canonical-depths}) is exact, address-native, and rounding-free.
\item \textbf{Centroid}: Calculate the arithmetic mean of all colors in the sub-cube and round each channel
value. This minimizes reconstruction error but is not an exact PTI address.
\end{enumerate}

The periodic-prefix convention is adopted as the natural representative for resolution
levels. It extends the address using the same recursive rule that defines PTI's depth
structure, relying solely on the address itself. In contrast, the centroid requires
referencing the underlying RGB values and is not a valid PTI address. The absence of an
exact centroid arises from the binary subdivision. At depth $d < 8$, each channel of a
sub-cube spans exactly $2^{8-d}$ integer values, which are always even, lacking a single
middle element. Therefore, the true channel-wise centroid always falls on a half-integer,
never an exact color.
The periodic-prefix convention's own tradeoff
(hierarchical consistency over Euclidean optimality) is discussed in
Section~\ref{sec:representative-tradeoffs}.

\begin{remark}[Truncation as Posterization] \label{rem:posterization} Truncation at canonical
depth $d$ in PTI yields results empirically equivalent to Photoshop's posterization at the
corresponding level. Unlike Photoshop, which allows an arbitrary selection of levels, PTI
truncation is limited to its predefined canonical depths ($d \in \{1,2,4,8\}$). Therefore, this
equivalence is valid only at the specific levels rather than across the full range of levels
that Photoshop supports. The potential computational advantages of this partial equivalence
remain to be investigated (Section~\ref{sec:future-work}). \end{remark}

\subsection{Rotating Addresses} \label{sec:rotating-addresses}

In the PTI addressing system, each color is represented by an 8-digit bit matrix, a
$3 \times 8$ binary array formed by the RGB channel bits. This structure supports a
cyclic left-shift operation on address digits, which has a clear geometric interpretation
in the RGB color space.

\subsubsection{Rotation Operation} \begin{definition}[Address Rotation] \label{def:rotation} Let $a = a_0a_1\ldots a_7$ be a PTI
address. The rotation of $a$ is the one-step cyclic left shift of its digits: $\Rot(a) =
a_1a_2\ldots a_7a_0$. $\Rot^k$ applies $\Rot$ exactly $k$ times in a row. The operation is defined for any valid PTI address, with no
constraints on digit distribution. \end{definition}

\begin{definition}[Rotation Orbit] \label{def:rotation-orbit} The rotation orbit of an address $a$
is the set of distinct addresses produced by repeated rotation: $\{a, \Rot(a), \Rot^2(a), \ldots\}$.
Since the address space is finite, this sequence eventually returns to $a$. The number of distinct
members, the orbit's period (Corollary~\ref{cor:rotation-period}), always divides 8. \end{definition}

\subsubsection{Rotation-Channel Correspondence} The rotation operation has a direct and exact
interpretation in RGB space. Recall that the PTI digit at position $i$ is $a_i = 4r_i +
2g_i + b_i$, where $r_i$, $g_i$, and $b_i$ are the $i$-th bits of the R, G, and B channel
values respectively. Rotating the address by one position shifts which column of the RGB
bit matrix appears at each digit position. This is precisely a one-bit circular left
shift applied independently to each channel.

Let $\ROL_k(X)$ denote the cyclic left rotation of an 8-bit channel value $X$
by $k$ bit positions. Bits shift left and the overflowing bits wrap to the right.
After 8 applications $\ROL_8(X) = X$ for any $X$.

\begin{theorem}[Rotation-Channel Correspondence] \label{thm:rotation-channel} For a color $C = (R,G,B)$ with PTI address
$a$, rotating $a$ by $k$ positions ($\Rot^k(a)$) results in the color $(\ROL_k(R),
\ROL_k(G), \ROL_k(B))$. \end{theorem}

A formal proof of the above theorem is provided in Appendix~\ref{app:proof-thm}.

\begin{corollary}[Rotation-Orbit Period] \label{cor:rotation-period} The rotation-orbit period of any PTI address divides 8.
Possible periods are 1, 2, 4, and 8, determined by
\[ \mathrm{lcm}\bigl(\mathrm{period}(R),\, \mathrm{period}(G),\, \mathrm{period}(B)\bigr), \]
where $\mathrm{period}(X)$ is the smallest $m \geq 1$ such that $\ROL_m(X) = X$. Period 1
occurs exactly for the eight anchors. \end{corollary}

This follows from Theorem~\ref{thm:rotation-channel}: since $\ROL_8(X) = X$ for any 8-bit
value $X$, the period of each channel divides 8, and therefore their LCM does as well. Eight
consecutive rotations return any address and its corresponding color to their original states.

The rotation operation presented here is the basis for palette derivation (Section~\ref{sec:applications}).

\section{Applications} \label{sec:applications}

\subsection{Palette Derivation in PTI}

Theorem~\ref{thm:rotation-channel} and Corollary~\ref{cor:rotation-period} establish that
every PTI address other than the eight anchors inherently contains a structured set of 2,
4, or 8 related colors, determined solely by the address itself. This
process, called palette derivation, extracts this set from a single seed address through
address rotation without requiring additional colors or coordinate conversions. The entire
computation remains within RGB, serving as an alternative to conventional methods for
formulating color palettes.

Colors within a rotation orbit share the same bit pool, with their RGB channel values
being cyclic permutations of an 8-bit sequence. Each subsequent color is generated by
shifting the starting position of these bits, maintaining consistency without creating or
destroying any bits.

Figure~\ref{fig:orbit-swatch} demonstrates the 8-color palette derived from a single seed
address through successive rotations:

\begin{figure}[H] \centering
\includegraphics[width=0.6\textwidth]{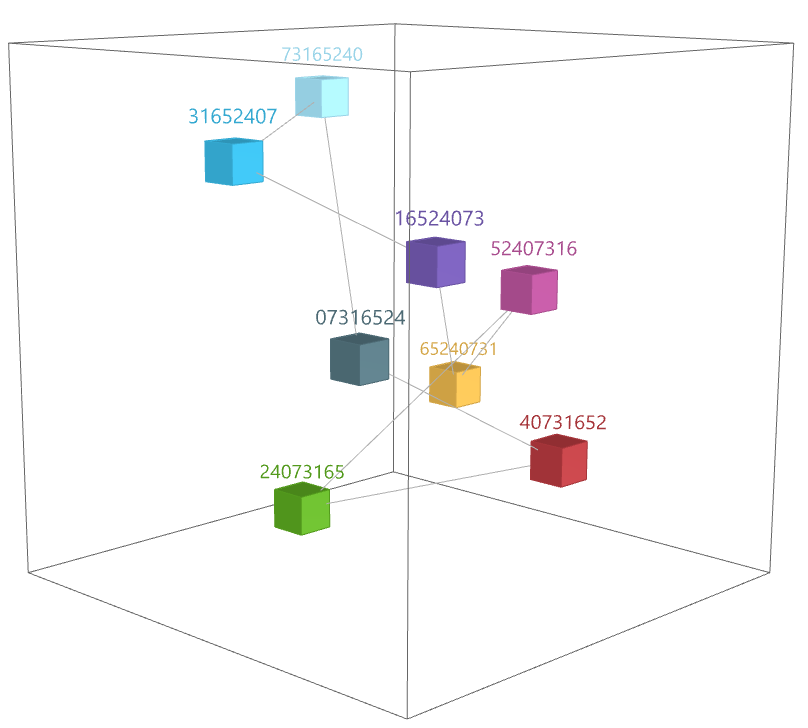}
\caption{8-color palette derived from seed address \texttt{07316524} through successive
rotations. An interactive version is available in the PTI web tools at
\url{https://metrigram.com/pti/orbit-explorer.html}.}
\label{fig:orbit-swatch}
\end{figure}

Each rotation orbit has a unique lexicographically smallest rotation, acting as its Lyndon
representative. For orbits with a period of 8, this minimal rotation is the Lyndon
word~\cite{lyndon1954burnside}; for shorter periods, it is derived from a repeated
Lyndon-word segment.

Future work on rotation-orbit taxonomy is discussed in Section~\ref{sec:future-work}.

\subsection{Color Explorer}

PTI's Color Explorer begins with a complete map of $\mathcal{C}_{24}$, the full color
space, at the family level, all 64 depth-2 families, arranged in an $8 \times 8$ grid
indexed by their leading two digits. Each color in $\mathcal{C}_{24}$ belongs to exactly
one cell. This is neither a color wheel nor a continuous picker. Exploration involves
selecting a family and drilling deeper rather than adjusting a continuous value.

\begin{figure}[H] \centering
\includegraphics[width=0.7\textwidth]{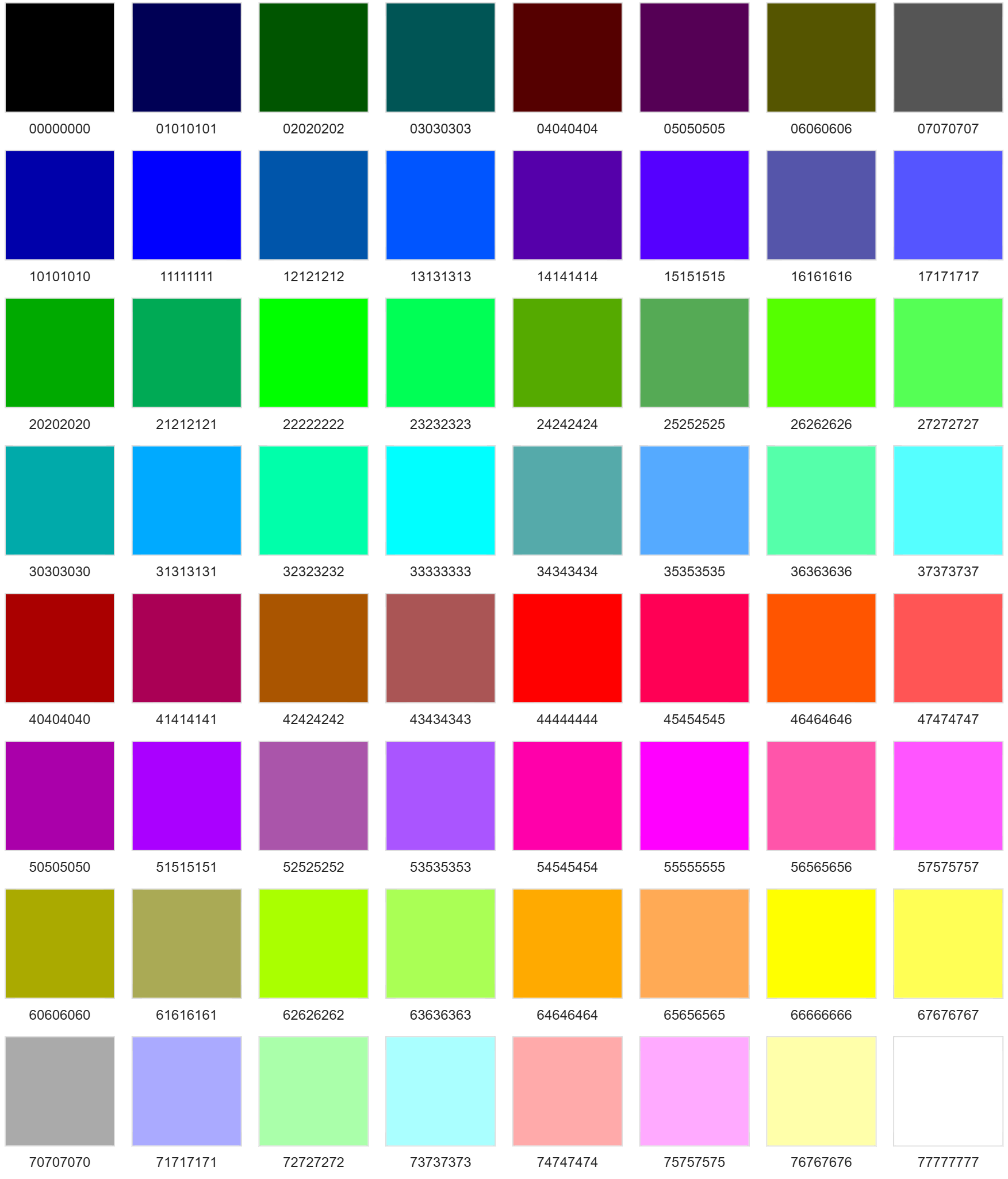}
\caption{All 64 depth-2 colors arranged in the $8\times8$ grid, showing the main diagonal
(anchors), anti-diagonal (complement pairs), and row coherence.}
\label{fig:depth2map} \end{figure}

The position of each family in the grid is given by $8a_0 + a_1$ for the address
$\overline{a_0 a_1}$, which reveals several structural properties. Figure~\ref{fig:depth2map}
illustrates these properties for all 64 depth-2 colors.

\subsubsection{Main Diagonal} The eight anchor colors appear at positions $9k$ (for $k = 0,
\ldots, 7$), since $8k + k = 9k$. These anchors partition the main diagonal.

\subsubsection{Anti-Diagonal} Addresses $\overline{a_0 a_1}$ with $a_0 + a_1 = 7$ occupy the
anti-diagonal at positions $7a_0 + 7$. These addresses pair complementary anchors, with
channel values in $\{85, 170\}$, corresponding to the vertices of a concentric inner cube.

\subsubsection{Row and Column Coherence} All eight addresses in row $a_0$ share the leading
digit and belong to the same depth-1 hue family, the row representing variations within one
primary anchor direction. Similarly, all eight addresses in column $a_1$ share the second
digit, the column representing that shared value across all eight families. Together, these
coherent rows and columns, along with the anchor diagonal and the complementary anti-diagonal,
follow directly from the digit-to-anchor arithmetic used above, requiring no additional
structure.

\textbf{Example}: Address $\overline{34}$ (Cyan-Red) occupies grid position $8 \times 3 +
4 = 28$. Its row contains all Cyan-family addresses: 30, \ldots, 37. Since $3 + 4 = 7$,
position $\overline{34}$ lies on the anti-diagonal, paired with its complement $\overline{43}$
(Red-Cyan).

\subsubsection{Recursive Exploration} Selecting a family reveals a new 64-cell grid focused on
that selection, narrowing the address by two more digits. This recursion continues through
depths 2, 4, 6, and 8, with depths 2, 4, and 8 canonical and depth 6 non-canonical. Colors
displayed at the canonical depths are exact periodic representatives, while colors displayed at
the non-canonical depth-6 step are expressed via an alternative representative, a direction
developed further in future work. Colors converge to the exact color only at depth 8. Each
selection also builds a compositional name, adding one anchor pair per level. An interactive
version of the Color Explorer is available in the PTI web tools at
\url{https://metrigram.com/pti/color-explorer.html}.

\subsection{Color Cataloging}

Every color naturally belongs to nested color families encoded in the address itself. PTI is
not a replacement for traditional color catalogs. It is an additional option, functioning as an
indexing and organization system for digital color data. Its primary strength lies in enabling
hierarchical grouping, browsing, and retrieval through simple address-prefix operations.

\begin{figure}[H] \centering
\includegraphics[width=\textwidth]{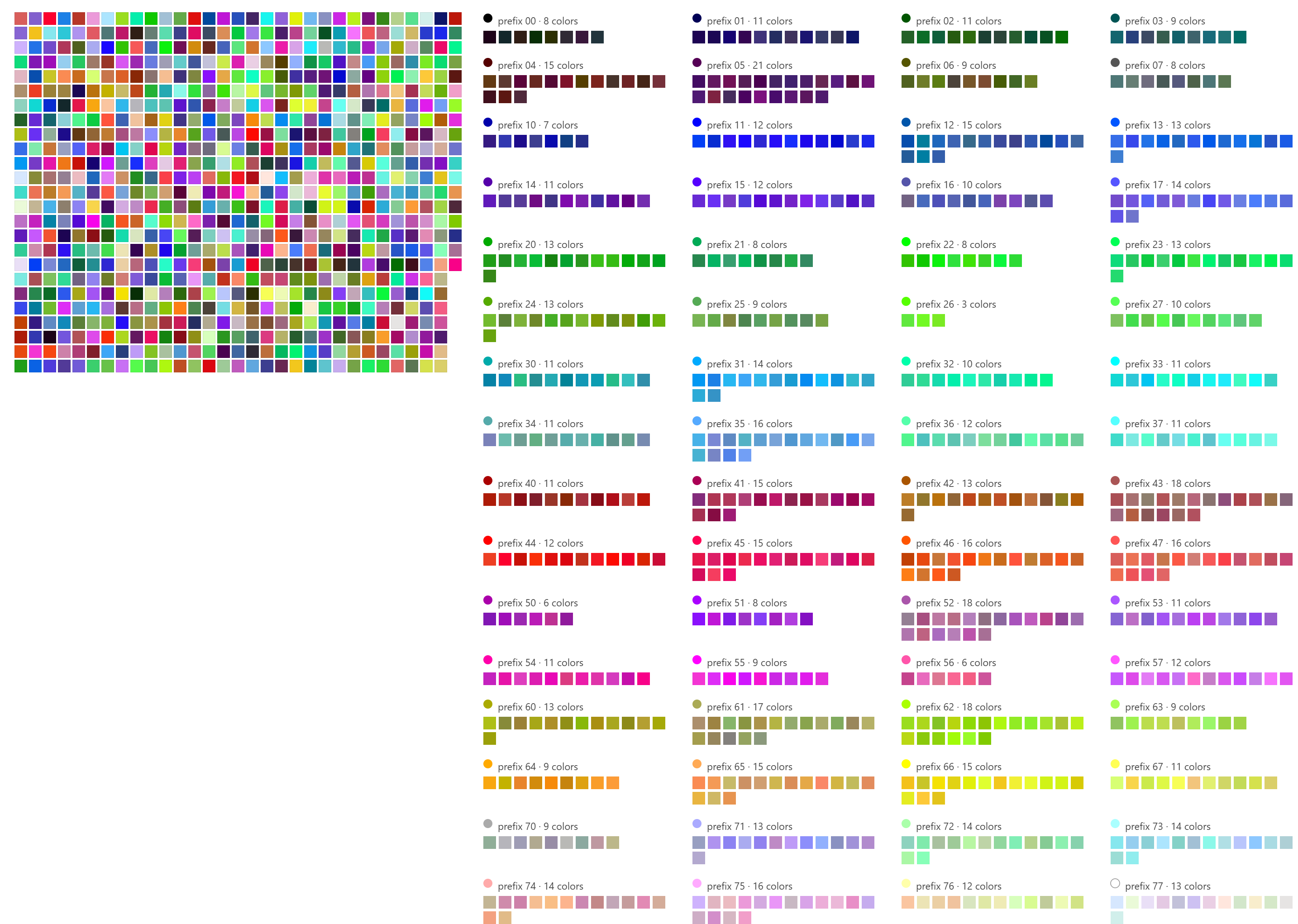}
\caption{A random catalog of 768 digital colors (left), organized by PTI depth-2 family
(right). Each populated family is labeled with its address prefix and color count, followed
by the individual colors collected in that family, requiring no perceptual computation or
manual sorting.}
\label{fig:catalog-grid}
\end{figure}

\section{Limitations} \label{sec:limitations}

\subsection{Not a Perceptual Space}

PTI is a re-encoding of the RGB cube and inherits all its limitations.

Its subdivision is geometrically uniform. Applications requiring perceptually uniform
similarity belong in CIELAB or a similar perceptual model, not in PTI address space.

\subsection{Canonical Representative Tradeoffs} \label{sec:representative-tradeoffs}

The canonical representative at each prefix depth is defined by address periodicity, not by
Euclidean proximity to the original color. This is a deliberate tradeoff. The
periodic-prefix convention prioritizes hierarchical consistency, the same rule that defines
PTI's own depth structure, over numerical optimality.

One consequence is that a coarser depth's representative can be closer, in Euclidean distance,
to the original color than a finer depth's representative. Across all 16,777,216 possible
colors, this occurs for 86,448 colors (approximately 0.52\%). The depth-2 representative is
strictly closer to the original color than the depth-4 representative. Within this group, eight
colors share the largest such gap: RGB $(79, 79, 79)$, PTI address \texttt{07007777}, is one of
them. Its depth-4 representative, \texttt{07000700}, decodes to $(68, 68, 68)$, at Euclidean
distance approximately 19.05 from the original color. Its depth-2 representative,
\texttt{07070707}, decodes to $(85, 85, 85)$, at distance approximately 10.39, achieving
roughly half the depth-4 representative's distance despite encoding only half as many of the
address's digits. A separate, further 268,288 colors (approximately 1.6\%) produce an exact tie
in Euclidean distance between their own depth-2 and depth-4 representatives.

The periodic-prefix representative remains canonical, the natural choice for
PTI's hierarchical structure. Still, there are some directions toward a possible alternative
representative for sub-cubes, which could be especially useful in cases like the one described
above, among others. These are deferred to future work.

\subsection{Compact Address Caveat}

At full depth the address is 8 characters versus 6 for hexadecimal. Compactness applies
only at reduced depths.

Additionally, while compact address notation offers syntactically terse encodings for
colors, there is a direct correlation between compactness and the number of available
colors and resolution. For example, the most compact address shorthand at depth $d=1$
can be represented with a single digit, but it provides a coarse resolution with only
eight possible colors at this depth. This is not an additional limitation. Compact address
notation \emph{is} the periodic representative itself (Remark~\ref{rem:compact}), so it
inherits exactly the tradeoffs discussed in Section~\ref{sec:representative-tradeoffs},
rather than introducing new ones.

In a broader context, PTI is not a standard encoding and requires implementation of
encode/decode routines in target systems. The upshot is that these routines are composed
of simple, closed-form bitwise operations.

\section{Future Work} \label{sec:future-work}

Several directions are identified for future work. PTI addresses support a formal query layer:
membership tests, class intersections, prefix-bounded enumeration, and direct address-level
constraints such as digit-value filters at specified positions are all derivable directly from
the address string, without decoding to RGB. This generality extends to specific structural
classes already identified in this paper, such as depth-defined periodic subsets and pandigital
rotation orbits. Building, mapping, naming, and researching the broader taxonomy of PTI
addresses is itself a future direction.

\begin{figure}[H] \centering
\includegraphics[width=0.6\textwidth]{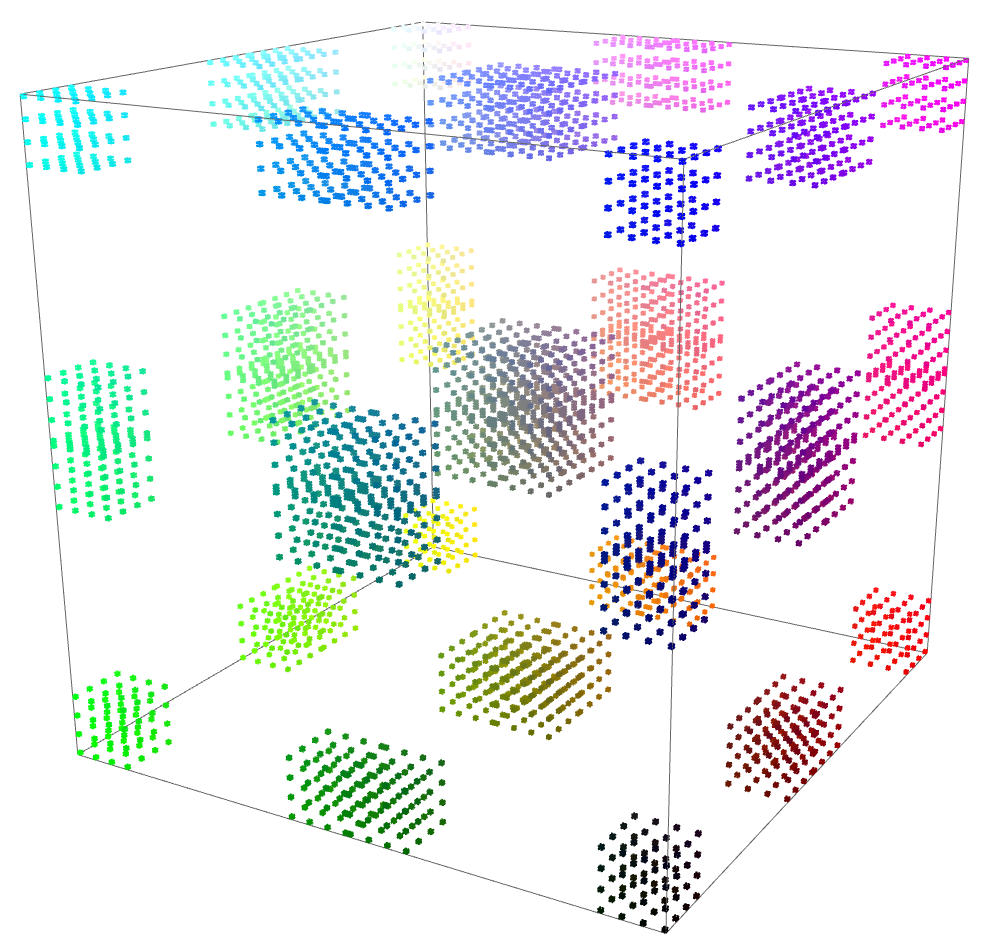}
\caption{A digit-value filter query across $\mathcal{C}_{24}$: fixing a subset of digit
positions while leaving the rest free selects a sparse scatter of sub-cubes distributed
throughout the RGB cube.}
\label{fig:queries-example}
\end{figure}

PTI's depth structure suggests a natural application to indexed color formats. Uniform 8-color
and 64-color palettes correspond to the canonical PTI depth-1 and depth-2 sets, respectively.
Furthermore, PTI offers a compact address notation for each color at these depths
(Remark~\ref{rem:compact}). Whether compact address notation at these depths could offer more
efficient image representation or compression is left as an open question for future
exploration.

Separately, the address rotation operation is not limited to PTI addresses. It has been found that
applying the same operation to hexadecimal color codes also derives native palettes.
Further formalization of this hex-rotation structure is left for future work.

This foundational specification outlines the core operations, applications, and properties
of the color addressing system. As PTI sees real-world use, additional extensions beyond these
directions are expected to emerge over time.

\section{Conclusion} \label{sec:conclusion}

This paper introduces PTI, a hierarchical symbolic addressing system for digital colors. The
primary contribution is the formalization of the persistent traversal identity within recursive
RGB cube subdivision. Instead of discarding the sequence of octant indices after locating a
color in an 8-level octree, it retains this sequence as the color's identifier.

PTI is positioned within the existing literature:

\begin{itemize}
\item Structurally, the complete PTI address corresponds to the Morton (Z-order)
code~\cite{morton1966} of the color, divided into consecutive 3-bit segments and interpreted
symbolically rather than as a scalar value.
\item Algorithmically, constructing the PTI address follows the octree traversal method
outlined by Gervautz and Purgathofer~\cite{gervautz1988simple}. This process aligns with
the structural correspondence above. At each level of the octree, the chosen octant index
constitutes the next 3-bit segment of the Morton code.
\item PTI does not propose a new hierarchical color subdivision technique. Rather, it suggests
treating the traversal path through this hierarchy as a named, persistent symbolic identifier.
\end{itemize}

The digit structure of PTI addresses allows certain color relationships to be directly
inferred or computed from the address itself. Grayscale colors are those that exclusively use
the digits 0 and 7 (Property~\ref{prop:grays}), and the additive complement of any color can
be derived by substituting each digit $d$ with $7 - d$ (Property~\ref{prop:complement}). The
rotation structure of PTI addresses leads to additional results, including a theorem
stating that rotating an address one position yields a specific color transformation
(Theorem~\ref{thm:rotation-channel}).

PTI is presented here as a foundational specification, with the aforementioned properties
and specifications established within this paper.

\newpage
\appendix

\section{Proofs} \label{sec:appendix-proofs}

This appendix provides full proofs for two results stated in the main text: the
correspondence between address rotation and per-channel bit rotation
(Theorem~\ref{thm:rotation-channel}), and the equivalence between PTI depth-4 periodic
representatives and CSS 3-digit shorthand hex (Theorem~\ref{thm:css-equiv}).

\subsection{Proof of Theorem~\ref{thm:rotation-channel}} \label{app:proof-thm}

\noindent\textit{Notation.} Write each channel value MSB-first in binary:
$R = \sum_{i=0}^{7} r_i \cdot 2^{7-i}$, and analogously for $G$ and $B$, with bits $r_i, g_i,
b_i \in \{0,1\}$. The PTI digit formula gives $a_i = 4r_i + 2g_i + b_i$. Each digit packages
one bit from each channel.

\begin{proof} Let $a' = \Rot^k(a)$, therefore $a'_i = a_{(i+k)\bmod 8}$ for each $i$.
Extracting the red-channel bit:
\[
  r'_i = \left\lfloor \frac{a'_i}{4} \right\rfloor
       = \left\lfloor \frac{a_{(i+k)\bmod 8}}{4} \right\rfloor
       = r_{(i+k)\bmod 8}.
\]
Reconstructing the red channel value:
\[
  R' = \sum_{i=0}^{7} r'_i \cdot 2^{7-i}
     = \sum_{i=0}^{7} r_{(i+k)\bmod 8} \cdot 2^{7-i}
     = \ROL_k(R).
\]
The identical calculation applies to green (extract via $g'_i = \lfloor(a'_i \bmod
4)/2\rfloor$) and blue (via $b'_i = a'_i \bmod 2$), giving $G' = \ROL_k(G)$ and $B' =
\ROL_k(B)$. \end{proof}

\subsection{Proof of Theorem~\ref{thm:css-equiv}} \label{app:proof-css-equiv}

\begin{proof} A PTI depth-4 periodic representative has the form $a_0 a_1 a_2 a_3 a_0 a_1
a_2 a_3$, where each digit is between 0 and 7. By the PTI decode formula, the red channel is
\[
R = \sum_{i=0}^{7} \left\lfloor \frac{a_i}{4} \right\rfloor \cdot 2^{7-i}.
\]
Since digits repeat with period 4 ($a_{i+4}$ equals $a_i$ for $i = 0, 1, 2, 3$), the floor
values in positions 4 through 7 equal those in positions 0 through 3. Therefore, the sum
splits into two sums, each running over $i$ from 0 to 3:
\[
R = \sum_{i=0}^{3} \left\lfloor \frac{a_i}{4} \right\rfloor \cdot 2^{7-i}
  + \sum_{i=0}^{3} \left\lfloor \frac{a_i}{4} \right\rfloor \cdot 2^{3-i}.
\]
Both sums share the same coefficients $\lfloor a_i/4 \rfloor$, therefore they factor together:
\begin{align*}
R &= \sum_{i=0}^{3} \left\lfloor \frac{a_i}{4} \right\rfloor \cdot \left(2^{7-i} + 2^{3-i}\right) \\
  &= \sum_{i=0}^{3} \left\lfloor \frac{a_i}{4} \right\rfloor \cdot 2^{3-i} \left(2^{4} + 2^{0}\right) \\
  &= 17 \sum_{i=0}^{3} \left\lfloor \frac{a_i}{4} \right\rfloor \cdot 2^{3-i} \\
  &= 17X,
\end{align*}
where $X = \sum_{i=0}^{3} \lfloor a_i/4 \rfloor \cdot 2^{3-i}$ is a 4-bit integer between 0
and 15. By identical reasoning, $G = 17Y$ and $B = 17Z$ for 4-bit integers $Y, Z \in
\{0,\ldots,15\}$. Therefore, every depth-4 periodic representative decodes to a color of the
form $(17X, 17Y, 17Z)$.

Conversely, the CSS shorthand \texttt{\#XYZ} expands to \texttt{\#XXYYZZ}, giving
channel values $17X$, $17Y$, $17Z$ for $X, Y, Z \in \{0,\ldots,15\}$ (hex digit labels,
unrelated to both the CIE XYZ color space and the generic channel placeholder $X$ used
earlier in $\ROL_k(X)$). The two sets are therefore identical. Both equal
$\bigl\{(17X,\,17Y,\,17Z) \mid X,Y,Z \in \{0,\ldots,15\}\bigr\}$, a bijection of $16^3 =
4096$ colors. \end{proof}

\clearpage
\printbibliography[title={References}]

\end{document}